\documentstyle[A4]{article}

\title{Non-perturbative solution of nonlinear Heisenberg equations}
\author{Ladislav Mi\v{s}ta, Jr.
\thanks{e-mail mista@optnw.upol.cz},
Radim Filip
\thanks{e-mail filip@optnw.upol.cz}
\\
Department of Optics,
Palack\'y University,\\
17. listopadu 50, 772 00 Olomouc,\\
Czech Republic}
\date{}

\begin{document}
\maketitle

\begin{abstract}
A new non-perturbative method of solution of the nonlinear Heisenberg
equations in finite-dimensional subspace is illustrated. The method,
being a counterpart of the traditional Schr\" odinger picture method,
is based on a finite operator expansion into the elementary processes.
It provides us with the insight into the nonlinear quantal interaction
from the different point of view. Thus one can investigate the
nonlinear system in both pictures of quantum mechanics.
\end{abstract}
\section{Introduction}

The use of laws of quantum mechanics in the description of nonlinear
systems confronts us with the qualitatively new difficulties. Namely,
to investigate their dynamics in the Heisenberg picture we have to solve
the nonlinear operator equations, a task which is highly nontrivial even for
the simplest systems. The difficulties are also encountered with in
the Schr\"odinger picture once we try to solve the Schr\"odinger
equation explicitly \cite{Andreev}. Since some nonlinear systems solvable
analytically in classical domain become insoluble when are quantized one can
suppose that they are simultaneous influence of intrinsic stochastic effects,
originating from the incompatibility of some observables, and
nonlinearity which make the behaviour of such systems very complex and thus
difficult to describe analytically.

The time evolution of quantum systems can be studied with the help of
widely used Schr\"odinger picture method based on the integration
of set of linear differential equations for components of a state
vector in the Fock basis \cite{Walls1},\cite{Nikitin}. Unfortunately,
the expansion into the Fock-state basis can be infinite for some states,
e.g. for coherent state, yielding the infinite set of these equations.
Because it is practically impossible to solve the infinite system of
the equations, the method provides us with exact solutions only for states
from some finite-dimensional subspace of the Hilbert state space.
On the other hand, it is advantageous sometimes to calculate the evolution
of particular observables in the framework of Heisenberg picture. The
motivation of the present paper is to find the operator analogue of the
Schr\" odinger picture method in the Heisenberg picture and to show the
equivalence and deep relationship between them.

As an illustrative example we consider here the simple system composed of two
harmonic oscillators which oscillate with frequencies $\omega$ and $2\omega$
and which are described by the annihilation (creation) operators
$\hat{a}_1(\hat{a}_1^{\dag})$ and $\hat{a}_2 (\hat{a}_2^{\dag})$ obeying
the standard boson type commutation rules
\begin{equation}\label{commutators}
[\hat{a}_i,\hat{a}_j^{\dag}]=\delta_{ij},\quad [\hat{a}_i,\hat{a}_j]=
[\hat{a}_i^{\dag},\hat{a}_j^{\dag}]=0,\quad i,j=1,2.
\end{equation}
Let the two oscillators interact nonlinearly according to the
following interaction Hamiltonian (the second-harmonic generation
process \cite{Crosignani})
\begin{equation}\label{ham}
\hat{H}=-\hbar\Gamma\hat{a}_1^{\dag 2}\hat{a}_2+\mbox{h.c.}\,,
\end{equation}
where $\Gamma$ denotes the nonlinear coupling constant; the symbol $\hbar$
is reduced Planck constant and h.c. stands for the Hermitian conjugate term.
Here and in the following we assume that the free evolution was eliminated
by the appropriate unitary transformation.

Employing the commutation rules (\ref{commutators}) one can
directly prove the existence of the following integral of motion
\begin{equation}\label{intmot}
\hat{N}=\hat{n}_{1}+2\hat{n}_{2},
\end{equation}
corresponding to the total energy of the system. Here the photon number
operator $\hat{n}_j$ of the $j$th oscillator , $j=1,2$ has been introduced.
The eigenvectors of the integral of motion (\ref{intmot}) then provide us
with the natural orthonormal and complete basis in which the expressions
have simple form. They are easy to find and have the form
\begin{equation}\label{basis}
\lbrace|N-2l,l\rangle,l=0,1,...,\Bigl[{\frac{N}{2}}\Bigr],N=0,1,
...\rbrace
\end{equation}
with orthonormality condition
\begin{equation}\label{ort}
\langle N-2l,l|M-2k,k\rangle=\delta_{NM}\delta_{lk}
\end{equation}
and the resolution of unity operator
\begin{equation}\label{res}
\sum_{N=0}^{\infty}\sum_{l=0}^{[\frac{N}{2}]}|N-2l,l\rangle\langle
N-2l,l|=\hat{1},
\end{equation}
where $|n_1,n_2\rangle$ is the Fock state having energy $\hbar\omega n_1+
2\hbar\omega n_2$; $N$ is the eigenvalue of (\ref{intmot}), $[N/2]$
represents the greatest integer less or equal to $N/2$ and $\delta_{lk}$
is the Kronecker symbol. The Hilbert state space of our system
can then be expressed as a direct sum
\begin{equation}\label{dirsum}
{\cal H}=\sum_{N=0}^{\infty}\oplus {\cal H}^{(N)}
\end{equation}
of the invariant $[N/2]+1$-dimensional subspaces ${\cal H}^{(N)}$ spanned
on the basis vectors $|N-2l,l\rangle,l=0,1,...,[N/2]$
corresponding to the fixed eigenvalue $N$. Using the standard
properties of the annihilation and creation operators of the harmonic
oscillator
\begin{eqnarray}\label{aprop}
\hat{a}_{i}|n_i\rangle&=&\sqrt{n_i}|n_i-1\rangle,\quad
\hat{a}_{i}^{\dag}|n_i\rangle=\sqrt{n_i+1}|n_i+1\rangle,
\quad i=1,2
\end{eqnarray}
and employing the condition (\ref{ort}) one can show that
the Hamiltonian (\ref{ham}) is represented by the following block
diagonal matrix
\begin{eqnarray}\label{hammatrix}
\langle N-2l,l|\hat{H}|M-2k,k\rangle&=&-\hbar[\Gamma\sqrt{(l+1)(N-2l)(N-2l-1)}
\delta_{k,l+1}\nonumber\\
&&+\Gamma^{\ast}\sqrt{l(N-2l+2)(N-2l+1)}\delta_{k,l-1}
]\delta_{NM},
\end{eqnarray}
where the symbol '$\ast$' represents the complex conjugation.
\section{Schr\" odinger picture}

Let us first recall the results obtained with the help of the Schr\"odiger
picture method when applied to our system. As is well-known the time
evolution of the state vector is governed by the Schr\"odinger
equation
\begin{equation}\label{Schreq}
i\hbar\frac{d}{dt}|\psi(t)\rangle=\hat{H}|\psi(t)\rangle,
\end{equation}
where Hamiltonian $\hat{H}$ is given in (\ref{ham}). Rewriting
(\ref{Schreq}) into the basis (\ref{basis}) with the help of
(\ref{res}) and (\ref{hammatrix}) we successively arrive at the
infinite number of sets of linear differential equations
\begin{eqnarray}\label{comp}
\frac{dC_{N,l}}{dt}&=&i\Gamma\sqrt{(l+1)(N-2l)(N-2l-1)}C_{N,l+1}
\nonumber\\
&&+i\Gamma^{\ast}\sqrt{l(N-2l+2)(N-2l+1)}C_{N,l-1}
\end{eqnarray}
for components $C_{N,l}\equiv\langle N-2l,l|\psi(t)\rangle$,
where $N=0,1,...$ and $l=0,1,...,[N/2]$. Assuming, however, the
initial state to be from the finite-dimensional subspace
\begin{equation}\label{finitsp}
{\cal H}_K=\sum_{N=0}^{K}\oplus{\cal H}^{(N)},
\end{equation}
it is sufficient to solve only $K+1$ such sets labelled by
eigenvalues $N=0,1,...,K$ each of them with $[N/2]+1$ equations.
Particularly, for states belonging to the subspaces ${\cal H}_{2}$,
the set (\ref{comp}) is of the form
\begin{equation}\label{loweq}
\frac{d C_{0,0}}{dt}=\frac{d C_{1,0}}{dt}=0,\quad
\frac{d C_{2,0}}{dt}=i\sqrt{2}\Gamma C_{2,1},\quad
\frac{d C_{2,1}}{dt}=i\sqrt{2}\Gamma^{\ast} C_{2,0}
\end{equation}
and can be solved analytically. An interesting result
is obtained assuming the system to be in the state $|0,1\rangle$
at the beginning of the interaction. The initial conditions for
the set (\ref{loweq}) are then $C_{0,0}(0)=C_{1,0}(0)=C_{2,0}(0)=0$,
$C_{2,1}(0)=1$ and the solution of (\ref{Schreq}) reads
\begin{equation}\label{sol}
|\psi(t)\rangle=\sum_{N=0}^{2}\sum_{l=0}^{[\frac{N}{2}]}
C_{N,l}(t)|N-2l,l\rangle=i\frac{\Gamma}{|\Gamma|}\sin(\sqrt{2}
|\Gamma|t)|2,0\rangle+\cos(\sqrt{2}|\Gamma|t)|0,1\rangle.
\end{equation}
Hence we obtain the following expressions for the mean number of
energy quanta in oscillators $1$ and $2$ in state (\ref{sol})
\begin{equation}\label{mean}
\langle\hat{n}_{1}(t)\rangle=2{\sin}^{2}(\sqrt{2}|\Gamma|t),\quad
\langle\hat{n}_{2}(t)\rangle={\cos}^{2}(\sqrt{2}|\Gamma|t).
\end{equation}
This non-classical oscillatory behaviour can be interpreted from
the point of view of Schr\"odiger picture as being a
manifestation of quantum interference effect (\ref{sol}).
\section{Heisenberg picture}

In this picture the operators $\hat{a}_1$ and $\hat{a}_2$ for the
system of interest evolve according to the Heisenberg equations of
motion
\begin{equation}\label{equation1}
i\hbar\frac{d\hat{a}_j}{dt}=[\hat{a}_j,\hat{H}],\quad j=1,2,
\end{equation}
which after substitution (\ref{ham}) into (\ref{equation1})
and application (\ref{commutators}) read
\begin{equation}\label{equation2}
\frac{d\hat{a}_1}{dt}=2i\Gamma\hat{a}_1^{\dag}\hat{a}_2,\quad
\frac{d\hat{a}_2}{dt}=i\Gamma^{\ast}\hat{a}_1^2.
\end{equation}
It is also well-known that the operators $\hat{a}_1(t)$ and
$\hat{a}_2(t)$ can be equivalently expressed as follows
\begin{equation}\label{pertsol}
\hat{a}_j(t)=\mbox{exp}(\frac{i}{\hbar}\hat{H}t)\hat{a}_j(0)
\mbox{exp}(-\frac{i}{\hbar}\hat{H}t)=\sum_{k=0}^{\infty}\frac{1}{k!}
\frac{d^k\hat{a}_j(0)}{dt^k}t^k,\quad j=1,2,
\end{equation}
where the exponential operators have been expanded and (\ref{equation1})
has been used repeatedly.

The power series on the right hand side (R.H.S.) of (\ref{pertsol}) is a
perturbative solution of the equations (\ref{equation2})  and provides us
with two important informations. As becomes clear from the following it is
advantageous to work in the normal ordering of the operators in which all
creation operators stand to the left from all annihilation operators. First,
the operator part of the solution (\ref{pertsol}), given by derivatives of
operators $\hat{a}_1(t)$ and $\hat{a}_2(t)$ at $t=0$, cannot contain
products of operators other than those of leading to the annihilation of
one energy quantum from the corresponding oscillator. This can be proved by
deriving the Heisenberg equations of motion (\ref{equation2}) and using
consequently the commutators (\ref{commutators}) to obtain the normally
ordered expressions. From now any such product of operators at $t=0$ is
called a {\it process} in the corresponding oscillator and the number
of operators in the product is called an {\it order} of the process.
Secondly, calculating the perturbative solution (\ref{pertsol}) to the
sufficiently high order and rearranging its terms appropriately, one can
see that the solution is of the form of finite sum of the processes
multiplied by various polynomials in $t$, which can constitute the first
few terms of power series of well-known functions (it can be verified
at least for the first few processes). This different point of view to
the standard perturbative solution \cite{Permon} is the core of our
{\it non-perturbative} method developed in the following text. Hence one
can surmise, that going to the infinity in the iterative procedure, the
solution of the Heisenberg equations of motion (\ref{equation2}) is of the
form of {\it infinite} sum of processes multiplied by some time dependent
functions
\begin{eqnarray}\label{sol1}
\hat{a}_1(t)&=&\hat{a}_1+f_1(t)\hat{a}_1^{\dag}\hat{a}_2+f_2(t)\hat{a}_1
^{\dag}\hat{a}_1^2+f_3(t)\hat{a}_2^{\dag}\hat{a}_1\hat{a}_2+...,
\end{eqnarray}
\begin{eqnarray}\label{sol2}
\hat{a}_2(t)&=&g_1(t)\hat{a}_2+g_2(t)\hat{a}_1^2+g_3(t)\hat{a}_1^{\dag}
\hat{a}_1\hat{a}_2+g_4(t)\hat{a}_2^{\dag}\hat{a}_2^2+...
\end{eqnarray}
where $\hat{a}_j\equiv\hat{a}_j(0)$, $j=1,2$. The functions $f_{j}$
and $g_{j}$ are called {\it amplitudes} of the corresponding
processes in the following text.

Substituting (\ref{sol1}) and (\ref{sol2}) into (\ref{equation2})
and comparing the coefficients related to the same process, the
amplitudes $f_j$ and $g_j$ can be determined as solutions of a system
of ordinary differential equations. For example, the equations for
amplitudes $f_1$ and $g_1$ together with the initial conditions
read
\begin{equation}\label{f1g1}
\frac{d}{dt}f_1(t)=2i\Gamma g_1(t),\quad \frac{d}{dt}g_1(t)=
i\Gamma^{\ast}f_1(t),\quad f_1(0)=0,\quad g_1(0)=1
\end{equation}
and have the following solutions
\begin{equation}\label{solf1g1}
f_1(t)=i\frac{\sqrt{2}\Gamma}{|\Gamma|}\sin(\sqrt{2}|\Gamma|t),\quad
g_1(t)=\cos(\sqrt{2}|\Gamma|t).
\end{equation}
Employing (\ref{sol1}) and (\ref{sol2}) the operators of the
number of the energy quanta in the oscillators $1$ and $2$ are of
the form
\begin{eqnarray}
\hat{n}_{1}(t)&=&\hat{a}_{1}^{\dag}\hat{a}_{1}+2\sin^{2}(\sqrt{2}
|\Gamma|t)\hat{a}_{2}^{\dag}\hat{a}_{2}+...,\label{n1}\\
\hat{n}_{2}(t)&=&\cos^{2}(\sqrt{2}|\Gamma|t)\hat{a}_{2}^{\dag}
\hat{a}_{2}+...,\label{n2}
\end{eqnarray}
where relations (\ref{commutators}) and (\ref{solf1g1}) have been
used. It is worth noting, that contrary to the R.H.S. of (\ref{n2})
the second term in (\ref{n1}) is the quantum contribution
originating from the commutator
$[\hat{a}_1,\hat{a}_{1}^{\dag}]=\hat{1}$. Considering as in the
previous section the input state to be $|0,1\rangle$ state and assuming
that the next terms represented by dots in (\ref{n1}) and (\ref{n2})
do not contribute, one obtains for the mean number of the energy quanta
in the oscillators the expressions
\begin{equation}\label{mean2}
\langle\hat{n}_1(t)\rangle=|f_1(t)|^2=2\sin^2(\sqrt{2}|\Gamma|t),
\quad
\langle\hat{n}_2(t)\rangle=|g_1(t)|^2=\cos^{2}(\sqrt{2}|\Gamma|t),
\end{equation}
which are identical with the results (\ref{mean}) obtained by means
of the Schr\" odinger picture method. Notice, that this derivation
illustrates not only the mathematical equivalence of both methods
but also their difference when one tries to distinguish between the
classical and quantum contributions.

Although one could look at the method just described as being a
satisfactory method, let us recall the reader, that its
conclusion (\ref{mean2}) rests on two crucial assumptions which were
not justified at all. First, we have assumed implicitly,
when deriving (\ref{f1g1}), that the higher order processes do
not affect the first order ones (cosequently we have obtaind the
finite set of differential equations for amplitudes $f_1$ and $g_1$).
Secondly, the mean numbers of energy quanta in state $|0,1\rangle$
given by (\ref{mean2}) have been derived under the assumption that
only explicitly given terms in (\ref{n1}) and (\ref{n2}) contribute.
To show that this is really the case, we have to formalise and
precise the Heisenberg picture method. This is done in the following
section.
\section{General method}

The previous section provides us with an illustrative example, how
one can treat the system (\ref{ham}) within the framework of Heisenberg
picture on the intuitive basis. In the present section we try to justify
the intuitive assumptions discussed above and to generalize this
treatment to the arbitrary finite-dimensional subspace. This can
be achieved by the suitable parametrization of the problem under
discussion. To that aim let us rewrite the expansions
(\ref{sol1}) and (\ref{sol2}) into the compact forms
\begin{eqnarray}\label{fijkl}
\hat{a}_{1}(t)&=&\sum_{i,j,k,l=0}^{\infty}f_{ijkl}(t)(\hat{a}_{1}
^{\dag})^{i}(\hat{a}_{2}^{\dag})^{j}\hat{a}_{1}^{k}\hat{a}_{2}^{l}
,\\
\hat{a}_{2}(t)&=&\sum_{o,p,r,s=0}^{\infty}g_{oprs}(t)
(\hat{a}_{1}^{\dag})^{o}(\hat{a}_{2}^{\dag})^{p}\hat{a}_{1}^{r}
\hat{a}_{2}^{s},\label{goprs}
\end{eqnarray}
where
\begin{equation}\label{cond1}
2l+k-2j-i=1
\end{equation}
and
\begin{equation}\label{cond2}
2s+r-2p-o=2
\end{equation}
holds. There are two facts which can make the convenient parametrization
easier to find. First, as in the Schr\"odinger picture we can
employ the existence of the integral of motion (\ref{intmot}).
Secondly, the discussion in the previous section indicates that
the order of the process is of importance. Therefore we put
\begin{equation}\label{MN1}
N=k+2l,\quad M=i+2j,\quad m=k+l,\quad R=i+j+k+l,
\end{equation}
for oscillator $1$ and similarly
\begin{equation}\label{MN2}
N=r+2s,\quad M=o+2p,\quad m=r+s,\quad R=o+p+r+s,
\end{equation}
for oscillator $2$. Thus each process is parametrized by
the parametres $N(M)$ (representing the amount of the annihilated
(created) energy in the process), $m$ (the total number of the
annihilated energy quanta) and $R$ (the order of the process).
Substituting (\ref{MN1}) into (\ref{fijkl}) and eliminating $N$
by means of (\ref{cond1}) we arrive at the following expansion
\begin{eqnarray}\label{f}
\hat{a}_{1}(t)&=&\sum_{M=0}^{\infty}\sum_{m=[\frac{M}{2}]+1}^{M+1}
\sum_{R=[\frac{M+1}{2}]+m}^{M+m}f_{MmR}(t)\nonumber\\
&&\times(\hat{a}_{1}^{\dag})^
{2R-M-2m}(\hat{a}_{2}^{\dag})^{M+m-R}\hat{a}_{1}^{2m-M-1}\hat{a}_{2}
^{M-m+1}.
\end{eqnarray}
In the same way we obtain
\begin{eqnarray}\label{g}
\hat{a}_{2}(t)&=&\sum_{M=0}^{\infty}\sum_{m=[\frac{M+1}{2}]+1}^{M+2}
\sum_{R=[\frac{M+1}{2}]+m}^{M+m}g_{MmR}(t)\nonumber\\
&&\times(\hat{a}_{1}^{\dag})^
{2R-M-2m}(\hat{a}_{2}^{\dag})^{M+m-R}\hat{a}_{1}^{2m-M-2}\hat{a}_{2}
^{M-m+2}.
\end{eqnarray}
Substituting (\ref{f}) and (\ref{g}) into the Heisenberg
equations of motion (\ref{equation2}), using the following
equation \cite{Permon}
\begin{equation}\label{newparam}
\hat{a}_{i}^{m}(\hat{a}_{i}^{\dag})^{n}=\sum_{j=0}^{min(m,n)}j!
{m\choose j}{n\choose j}(\hat{a}_{i}^{\dag})^{n-j}
\hat{a}_{i}^{m-j},\quad i=1,2,
\end{equation}
which can be proved easily using the commutation rules
(\ref{commutators}), and comparing the expressions corresponding
to the same process, we obtain the following {\it infinite} set of
differential equations for amplitudes $f_{MmR}, M=0,1,...$;
$m=[\frac{M}{2}]+1,...,M+1$; $R=[\frac{M+1}{2}]+m,...,M+m$ and
$g_{MmR}, M=0,1,...$; $m=[\frac{M+1}{2}]+1,...,M+2$; $R=[\frac{M+1}{2}]+m
,...,M+m$:
\begin{eqnarray}\label{derivf}
\frac{d}{dt}f_{MmR}(t)&=&2i\Gamma\sum_{M_{1},M_{2}=0}^{\infty}
\sum_{m_{1}=[\frac{M_{1}}{2}]+1}^{M_{1}+1}\sum_{m_{2}=[\frac{M_{2}+1}
{2}]+1}^{M_{2}+2}\sum_{R_{1}=[\frac{M_{1}+1}{2}]+m_{1}}^{M_{1}+
m_{1}}\sum_{R_{2}=[\frac{M_{2}+1}{2}]+m_{2}}^{M_{2}+m_{2}}
\nonumber\\
&&\times {2R_{1}-M_{1}-2m_{1}\choose s_1}{2R_{2}-M_{2}-2m_{2}\choose s_1}
{M_{1}+m_{1}-R_{1}\choose s_2}\nonumber\\
&&\times{M_{2}+m_{2}-R_{2}\choose s_2}s_{1}!s_{2}!
f_{M_{1}m_{1}R_{1}}^{\ast}(t)g_{M_{2}m_{2}R_{2}}(t),
\end{eqnarray}
\begin{eqnarray}\label{derivg}
\frac{d}{dt}g_{MmR}(t)&=&i\Gamma^{\ast}\sum_{M_{1},M_{2}=0}^{\infty}
\sum_{m_{1}=[\frac{M_{1}}{2}]+1}^{M_{1}+1}
\sum_{m_{2}=[\frac{M_{2}}{2}]+1}^{M_{2}+1}\sum_{R_{1}=
[\frac{M_{1}+1}{2}]+m_{1}}^{M_{1}+m_{1}}\sum_{R_{2}=
[\frac{M_{2}+1}{2}]+m_{2}}^{M_{2}+m_{2}}\nonumber\\
&&\times {2m_{1}-M_{1}-1\choose s_1}{2R_{2}-M_{2}-2m_{2}\choose s_1}
{M_{1}-m_{1}+1\choose s_2}\nonumber\\
&&\times{M_{2}+m_{2}-R_{2}\choose s_2}s_{1}!s_{2}!
f_{M_{1}m_{1}R_{1}}(t)f_{M_{2}m_{2}R_{2}}(t),
\end{eqnarray}
where
\begin{eqnarray}\label{sR1}
R_{2}&=&R+R_{1}-2m-2m_{1}+2m_{2},\nonumber\\
s_{1}&=&M-M_{1}-M_{2}-2m-2m_{1}+2m_{2}+2R_{1}-1,\nonumber\\
s_{2}&=&M_{1}+M_{2}-M-m-m_{1}+m_{2}+R-R_{2}+1
\end{eqnarray}
for equation (\ref{derivf}) and
\begin{eqnarray}\label{sR2}
R_{2}&=&R-R_{1}-2m+2m_{1}+2m_{2},\nonumber\\
s_{1}&=&M-M_{1}-M_{2}+2m-2m_{1}-2m_{2}-2R+2R_{1}+2R_{2},\nonumber\\
s_{2}&=&M_{1}+M_{2}-M-m+m_{1}+m_{2}+R-R_{1}-R_{2}
\end{eqnarray}
for equation (\ref{derivg}). The initial conditions for equations
(\ref{derivf}) and (\ref{derivg}) are $f_{011}(0)=g_{011}(0)=1,
f_{MmR}(0)=g_{MmR}(0)=0$ in all other cases.

The structure of R.H.S. of (\ref{derivf}) reveals that there is a
nonzero contribution to the R.H.S. only if the following
inequalities hold simultaneously
\begin{eqnarray}\label{ineqf}
2R_{1}-M_{1}-2m_{1}&\ge& s_{1},\quad 2R_{2}-M_{2}-2m_{2}\ge s_{1},
\nonumber\\
M_{1}+m_{1}-R_{1}&\ge& s_{2},\quad M_{2}+m_{2}-R_{2}\ge s_{2}.
\end{eqnarray}
Combining this with (\ref{sR1}) one finally obtains
\begin{equation}\label{ineq1}
M_{1}\le M-1,\quad M_{2}\le M-1.
\end{equation}
Analogously, R.H.S. of (\ref{derivg}) contains nonzero contribution
only if
\begin{eqnarray}\label{ineqg}
2m_{1}-M_{1}-1&\ge& s_{1},\quad 2R_{2}-M_{2}-2m_{2}\ge s_{1},
\nonumber\\
M_{1}-m_{1}+1&\ge& s_{2},\quad M_{2}+m_{2}-R_{2}\ge s_{2},
\end{eqnarray}
hold simultaneously. Consequently, the R.H.S. of (\ref{derivg}) cannot
contain amplitudes other than those for which
\begin{equation}\label{ineq2}
M_{1}\le M,\quad M_{2}\le M+1.
\end{equation}
From the above inequalities (\ref{ineq1}) and (\ref{ineq2}) follows
that for fixed $M$ we have only {\it finite} set of equations
(\ref{derivf}), (\ref{derivg}) for amplitudes $f_{M'm'R'},
M'=0,1,...,M$; $m'=[\frac{M'}{2}]+1,...,M'+1$; $R'=[\frac{M'+1}{2}]+m'
,...,M'+m'$ and $g_{M''m''R''}, M''=0,1,...,M-1$; $m''=[\frac{M''+1}{2}]
+1,...,M''+2$; $R''=[\frac{M''+1}{2}]+m'',...,M''+m''$.
In other words, the process in the oscillator 1(2) parametrized by
$M'>M(M''>M-1)$ does not affect the processes in the oscillator
1(2) for which $M'\le M(M''\le M-1)$, as we wanted to prove.

The discussion of the structure of the set of equations
(\ref{derivf}), (\ref{derivg}) can go even further.
Since the amplitudes $f_{M'm'R'}, M'<M$  and $g_{M''m''R''},
M''<M-1$ can be calculated solving the set of equations
(\ref{derivf}), (\ref{derivg}) corresponding to $M-1$,
in fact only the amplitudes $f_{Mm'R'}, m'=[\frac{M}{2}]+1,...,M+1$;
$R'=[\frac{M+1}{2}]+m',...,M+m'$ and $g_{M-1m''R''},
m''=[\frac{M}{2}]+1,...,M+1$; $R''=[\frac{M}{2}]+m'',...,M+m''-1$
are mutually coupled. The amplitudes $f_{M'm'R'}, M'<M$  and
$g_{M''m''R''}, M''<M-1$ then play the role of known coefficients
and source terms and the set of differential equations
corresponding to $M$ is linear. Moreover, taking $M$ and $m$ fixed,
substituting (\ref{sR1}) into (\ref{ineqf}) and putting $M_{1}=M_{2}=M-1$
one obtains the following equality
\begin{equation}\label{eq1}
m_{2}=m.
\end{equation}
Since the same equality can be proved substituting (\ref{sR2})
into (\ref{ineqg}) and putting $M_{1}=M, M_{2}=M+1$, one can conclude
that only amplitudes $f_{MmR'}, R'=[\frac{M+1}{2}]+m,...,M+m$ and $g_{M-1mR''},
R''=[\frac{M}{2}]+m,...,M+m-1$ are coupled. Hence for
given $M$ and $m$ one has to solve the set of $M+1$ differential
equations (\ref{derivf}) and (\ref{derivg}).

Before going further let us notice that since the infinite series (\ref{f})
and (\ref{g}) with amplitudes being the solutions of the equations
(\ref{derivf}) and (\ref{derivg}) satisfy the Heisenberg equations of
motion (\ref{equation2}) identically, the operators $\hat{a}_{1}(t)$
and $\hat{a}_{2}(t)$ preserve the commutation rules (\ref{commutators}).

Someone still could object that our method cannot be used in
practice since we are not able to calculate all the amplitudes.
This difficulty is, however, overcome if one realizes the following
fact. Calculating the matrix element of the process corresponding
to parameters $M$, $m$ and $R$ for oscillator $1$,
\begin{eqnarray}\label{matrix}
\lefteqn{\langle N_{1}-2l_{1},l_{1}|(\hat{a}_{1}^{\dag})^{2R-M-2m}
(\hat{a}_{2}^{\dag})^{M+m-R}\hat{a}_{1}^{2m-M-1}\hat{a}_{2}^{M-m+1}
|N_{2}-2l_{2},l_{2}\rangle}\nonumber\\
&&=\sqrt{\frac{l_{1}!l_{2}!(N_{1}-2l_{1})!(N_{2}-2l_{2})!}
{(N_{1}-2l_{1}+M+2m-2R)!(N_{2}-2l_{2}+M-2m+1)!}}\nonumber\\
&&\times\frac{\delta_{N_{1}-2l_{1}+4m,N_{2}-2l_{2}+2R+1}\delta_
{l_{1}+R+1,l_{2}+2m}}{\sqrt{(l_{1}-M-m+R)!(l_{2}-M+m-1)!}},
\end{eqnarray}
where the formulas (\ref{ort}) and (\ref{aprop}) have been used, it is
evident that it does not vanish only if the following inequalities are
satisfied simultaneously
\begin{eqnarray}\label{ineq3}
N_{1}-2l_{1}&\ge& 2R-M-2m,\quad 2l_{1}\ge 2M+2m-2R,
\nonumber\\
N_{2}-2l_{2}&\ge& 2m-M-1,\quad 2l_{2}\ge 2M-2m+2.
\end{eqnarray}
Hence
\begin{equation}\label{ineq4}
N_{1}\ge M,\quad N_{2}\ge M+1.
\end{equation}
Repeating the same discussion for the same matrix element of the
process in oscillator $2$ characterized by the parameters $M$, $m$
and $R$ one arrives at
\begin{equation}\label{ineq5}
N_{1}\ge M,\quad N_{2}\ge M+2.
\end{equation}
The inequalities (\ref{ineq4}) and (\ref{ineq5}) can be
interpreted as follows. Restricting ourselves to the
finite-dimensional subspace ${\cal H}_{K}$ of the whole Hilbert
space (\ref{dirsum}) only processes in oscillator $1(2)$ for
which $M'=0,1,...,K-1(M''=0,1,...,K-2)$ are represented by
nonzero matrix. In other words, the time evolution of the
operators $\hat{a}_{1}(t)$ and $\hat{a}_{2}(t)$ on the subspace
${\cal H}_{K}$ is known once the amplitudes $f_{M'm'R'},
M'=0,1,...,K-1$; $m'=[\frac{M'}{2}]+1,...,M'+1$; $R'=[\frac{M'+1}{2}]+m'
,...,M'+m'$ and $g_{M''m''R''}, M''=0,1,...,K-2$; $m''=
[\frac{M''+1}{2}]+1,...,M''+2$; $R''=[\frac{M''+1}{2}]+m'',...,M''+m''$
are determined. This requires sequential solution of $[\frac{M}{2}]+1$
sets of $M=0,1,...,K$ differential equations (\ref{derivf}) and
(\ref{derivg}). Since the series (\ref{f}) and (\ref{g}) are terminated
naturally when considering only finite-dimensional subspace ${\cal H}_{K}$,
a natural question arises whether the last-named amplitudes determine not
only the evolution of $\hat{a}_{1}(t)$ and $\hat{a}_{2}(t)$ but also the
evolution of any operator on the subspace ${\cal H}_{K}$. Now we will
prove that this is really the case.

It is well known that any operator at time $t$ on the space ${\cal H}$
can be expressed as a sum of the following products
$(\hat{a}_{1}^{\dag})^{i}(t)(\hat{a}_{2}^{\dag})^{j}(t)\hat{a}_{1}^{k}
(t)\hat{a}_{2}^{l}(t)$. Hence, it is sufficient to prove the statement
for these products only. The commutation rules (\ref{commutators})
enable us to show that
\begin{eqnarray}\label{Npr1}
\lefteqn{\hat{N}(\hat{a}_{1}^{\dag})^{2R-M-2m}(\hat{a}_{2}^{\dag})^{M+m-R}
\hat{a}_{1}^{2m-M-1}\hat{a}_{2}^{M-m+1}|N-2l,l\rangle}
\nonumber\\
&&=(N-1)(\hat{a}_{1}^{\dag})^{2R-M-2m}(\hat{a}_{2}^{\dag})^{M+m-R}
\hat{a}_{1}^{2m-M-1}\hat{a}_{2}^{M-m+1}|N-2l,l\rangle
\end{eqnarray}
for oscillator $1$ and similarly
\begin{eqnarray}\label{Npr2}
\lefteqn{\hat{N}(\hat{a}_{1}^{\dag})^{2R-M-2m}(\hat{a}_{2}^{\dag})^{M+m-R}
\hat{a}_{1}^{2m-M-2}\hat{a}_{2}^{M-m+2}|N-2l,l\rangle}
\nonumber\\
&&=(N-2)(\hat{a}_{1}^{\dag})^{2R-M-2m}(\hat{a}_{2}^{\dag})^{M+m-R}
\hat{a}_{1}^{2m-M-2}\hat{a}_{2}^{M-m+2}|N-2l,l\rangle
\end{eqnarray}
for oscillator $2$. Consquently,
\begin{eqnarray}\label{Nprt}
\hat{N}\hat{a}_{j}(t)|N-2l,l\rangle&=&(N-j)\hat{a}_{j}(t)|N-2l,
l\rangle,\quad j=1,2,
\end{eqnarray}
as one can verify, using (\ref{f}) and (\ref{g}). From that it
follows that $\hat{a}_{j}(t){\cal H}_{K}\subset{\cal H}_{K-j}$,
$j=1,2$. Since the first annihilation (creation) operator to the
right (left) in the matrix elements
\begin{equation}\label{mom}
\langle N_{1}-2l_{1},l_{1}|(\hat{a}_{1}^{\dag})^{i}(t)
(\hat{a}_{2}^{\dag})^{j}(t)\hat{a}_{1}^{k}(t)\hat{a}_{2}^{l}(t)
|N_{2}-2l_{2},l_{2}\rangle,\quad N_{1},N_{2}=0,1,...,K
\end{equation}
transforms the basis vector to the right (left) into the subspace
embeded into ${\cal H}_{K}$, the series (\ref{f}) and (\ref{g}) for
the following annihilation (creation) operators must terminate even
further than those for the first annihilation (creation) operator.
Therefore no other amplitude except for those mentioned
above can appear in the expression (\ref{mom}). This is what we
wanted to prove.

There is one more point connected with the previous discussion
which should be clarified here. Namely, one could think about the
finite series for $\hat{a}_{1}(t)$ and $\hat{a}_{2}(t)$ on the
${\cal H}_{K}$ as an approximate operator solutions of
(\ref{equation2}). The following special example disproves the idea.

Let us consider the following finite series (describing
correctly the time evolution on the subspace ${\cal H}_{2}$)
\begin{eqnarray}
\hat{a}_{1}(t)&=&f_{011}(t)\hat{a}_{1}+f_{112}(t)\hat{a}_{1}^
{\dag}\hat{a}_{2}+f_{123}(t)\hat{a}_{1}^{\dag}\hat{a}_{1}^{2},
\label{a1}\\
\hat{a}_{2}(t)&=&g_{011}(t)\hat{a}_{2}+g_{022}(t)\hat{a}_{1}^{2}.
\label{a2}
\end{eqnarray}
The amplitudes in (\ref{a1}) and (\ref{a2}) are solutions of the
set of equations
\begin{eqnarray}\label{H1}
\frac{d}{dt}f_{011}(t)&=&0,\nonumber\\
\frac{d}{dt}f_{112}(t)&=&2i\Gamma f_{011}^{\ast}(t)g_{011}(t),\quad
\frac{d}{dt}g_{011}(t)=i\Gamma^{\ast}f_{011}(t)f_{112}(t),\nonumber\\
\frac{d}{dt}f_{123}(t)&=&2i\Gamma f_{011}^{\ast}(t)g_{022}(t),\quad
\frac{d}{dt}g_{022}(t)=i\Gamma^{\ast}f_{011}(t)(f_{011}(t)+f_{123}(t)),
\nonumber\\
\end{eqnarray}
with the initial conditions $f_{011}(0)=g_{011}(0)=1$, $f_{112}(0)
=f_{123}(0)=g_{022}(0)=0$. The use of standard methods then yields
\begin{eqnarray}\label{solH2}
f_{011}(t)&=&1,\nonumber\\
f_{112}(t)&=&i\frac{\sqrt{2}\Gamma}{|\Gamma|}\sin(\sqrt{2}|\Gamma|t),
\quad g_{011}(t)=\cos(\sqrt{2}|\Gamma|t),\nonumber\\
f_{123}(t)&=&\cos(\sqrt{2}|\Gamma|t)-1,\quad g_{022}(t)=
i\frac{\Gamma^{\ast}}{\sqrt{2}|\Gamma|}\sin(\sqrt{2}|\Gamma|t).
\end{eqnarray}
Now, substituting (\ref{solH2}) into (\ref{a2}) one arrives at
\begin{equation}\label{comut2}
[\hat{a}_{2}(t),\hat{a}_{2}^{\dag}(t)]=\hat{1}+2\sin^{2}(\sqrt{2}
|\Gamma|t)\hat{a}_{1}^{\dag}\hat{a}_{1}\ne \hat{1}.
\end{equation}
In the same way we can prove that $[\hat{a}_{1}(t),\hat{a}_{1}^{\dag}
(t)]\ne \hat{1}$. This illustrates that the commutation rules are not
preserved for the finite series (\ref{a1}) and (\ref{a2}). One can
expect this, since the series (\ref{a1}) and (\ref{a2}) represent a
correct solution only on the subspace ${\cal H}_{2}$. Thus only the
complete solution (\ref{f}) and (\ref{g}) involving all the processes
preserves the commutation rules in the whole Hilbert space.
\section{Conclusion}

On the simple example we illustrate how to solve the nonlinear Heisenberg
equations of motion on the finite-dimensional subspace using the finite
expansion of annihilation operators into the sum of the elementary processes.
The idea of the method is not restricted to this example
and provides us with recipe how to treat other nonlinear
interactions. The time evolution of any operator on the subspace is then
governed by a finite number of the c-number differential equations for
amplitudes. Due to the hierarchy of the processes the equations split
into several sets which can be solved step by step. Thus the problem of
solution of the q-number Heisenberg equations is transformed into the
finding of solution of the linear c-number differential equations, which
can be handled numerically. It provides a nice interpretation and deeper
insight into what happens in the course of the nonlinear quantal interaction
in the language of elementary processes. It also enables us to identify the
non-classical contributions. This instructive interpretation cannot be
obtained within the framework of the Schr\"odinger picture.
\section{Acknowledgments}

We would like to thank Prof. J. Pe\v{r}ina for useful advices.
This work was supported by Grant LN00A015 of Czech Ministry of
Education.

\end{document}